\documentclass[a4paper,11pt]{article}
\pdfoutput=1

\usepackage{soul}
\usepackage{jcappub} % for details on the use of the package, please % see the JHEP-author-manual
\usepackage{tikz}
\usepackage{xcolor}

\usepackage[utf8]{inputenc}
\usepackage{amsmath, amsthm, mathtools}
\usepackage{braket}

\usepackage{tensor}
\usepackage{verbatim}
\usepackage{amsmath,amsfonts,amssymb}

\usepackage{fontawesome}

\usepackage{graphicx}
\usepackage{adjustbox}
\usepackage{tabularx}
\usepackage{tensor}
\usepackage{amsbsy}
\usepackage{braket}
\usepackage{slashed}
\usepackage{enumitem}
\hypersetup{%
,urlcolor=blue
,citecolor=blue
,linkcolor=blue
}

\usetikzlibrary{shapes.geometric, arrows}

\tikzstyle{startstop} = [rectangle, rounded corners, minimum width=3cm, minimum height=1cm,text centered, draw=black, fill=red!30]
\tikzstyle{process} = [rectangle, minimum width=3cm, minimum height=1cm, text centered, draw=black, fill=orange!30]
\tikzstyle{decision} = [diamond, minimum width=3cm, minimum height=1cm, text centered, draw=black, fill=green!30]
\tikzstyle{arrow} = [thick,->,>=stealth]

\usepackage[T1]{fontenc} % if needed

 \usepackage{braket}

\title{\boldmath AC Currents from Gravitational Waves in Plasma Flows}

\author[]{J. I. McDonald}
\emailAdd{jamie.mcdonald@manchester.ac.uk}

 \affiliation[]{
Department of Physics and Astronomy\\
University of Manchester\\
 Manchester, M13 9PL, UK}

\abstract{It is well-known that gravitational waves can induce electromagnetic perturbations in magnetised plasmas, with production occurring via the direct coupling of gravitational waves to the background magnetic field: this is the so-called Gertsenshtein effect.  In  this short work, we consider the direct gravitational perturbations of charge carriers via their minimal coupling to gravity in a collisionless plasma. We find that for isotropic plasmas, no secondary plasma perturbations are generated. However, when an anisotropy is introduced in the form of a background plasma current, we find that gravitational waves can induce a secondary current. For a constant DC background current, the secondary current inherits the AC frequency of the gravitational waves. It will certainly be interesting to investigate this effect in astrophysical plasmas in future work as well as its wider phenomenological consequences.   }

\bigskip

\tikzstyle{startstop} = [rectangle, rounded corners, minimum width=3cm, minimum height=1cm, text centered, draw=black, fill=red!30]
\tikzstyle{process} = [rectangle, minimum width=4cm, minimum height=1cm, text centered, draw=black, fill=orange!30]
\tikzstyle{stop} = [rectangle, rounded corners, minimum width=3cm, minimum height=1cm, text centered, draw=black, fill=yellow!30] % Stop node style
\tikzstyle{arrow} = [thick,->,>=stealth]

\begin{document}

\notoc

\maketitle
\flushbottom

\newpage

% \section*{}

 \section{Introduction}

 The propagation of gravitational waves in media, including plasmas \cite{Macedo:1983wcr}, has been the subject of much interest over the years with particular focus on the well-known Gertsenshtein effect  \cite{Gert61,Palessandro:2023tee} in which gravitational waves are converted into electromagnetic perturbations in the presence of a magnetic field. This effect has been used extensively in plasmas to constrain the existence of gravitational wave backgrounds using the magnetic fields of galaxies \cite{Lella:2024dus,Ito:2023nkq}, planetary magnetospheres \cite{Liu:2023mll} neutron stars \cite{Ito:2023fcr,McDonald:2024nxj,Dandoy:2024oqg} and the CMB \cite{Domcke:2020yzq}, as well as more directly via existing laboratory setups where a magnetic field is present (see, e.g., \cite{Ejlli:2019bqj,Berlin:2021txa,Domcke:2022rgu}). Let us now re-examine gravitational waves in plasmas. 

 Consider linearised metric perturbations $h_{\mu \nu}$ about flat space, so that the full metric $g_{\mu \nu}$ reads
 \begin{equation}
     g_{\mu \nu } = \eta_{\mu \nu} + h_{\mu \nu},
 \end{equation}
where $\eta_{\mu \nu}$ is the Minkowski metric, then it is well-known that the effective interaction vertex between the gravitational perturbations and matter fields is given by
\begin{equation}\label{eq:Lint}
    \mathcal{L}_{\rm int} =  h_{\mu \nu} T^{\mu \nu},
\end{equation}
where $T_{\mu \nu}$ is the energy-momentum tensor of matter. If one considers gauge transformations on $h_{\mu \nu}$ of the form
 \begin{equation}
     h_{\mu \nu }  \rightarrow h_{\mu \nu} +  \frac{1}{2}\left( \partial_{ \mu} \xi_{\nu} +  \partial_{ \nu} \xi_{\mu} \right) ,
 \end{equation}
 it then follows that 
 \begin{equation}
     \mathcal{L}_{\rm int} \rightarrow     h_{\mu \nu} T^{\mu \nu} + \partial_{ \mu} \xi_{\nu} T^{\mu \nu}  =  h_{\mu \nu} T^{\mu \nu}  -  \xi_{\nu} \partial_\mu T^{\mu \nu}, 
 \end{equation}
 where, after an integration by parts, the second equality holds up to a total derivative.  We therefore see that the gauge invariance of the interaction vertex is ensured automatically by energy-momentum conservation:
 \begin{equation}
     \partial_{\mu } T^{\mu \nu} = 0 .
  \end{equation}
In the case of a plasma, the full energy momentum tensor is made of two components
\begin{equation}
 T^{\mu \nu}  = T_{\gamma}^{\mu \nu} + T_{e}^{\mu \nu}\, ,
\end{equation}
where 
\begin{equation}\label{eq:EMTensor}
T^{\mu \nu}_{\gamma} =  F^{\mu \alpha} F^{\nu}{}_{\alpha} - \frac{1}{4} \eta^{\mu \nu} F_{\alpha \beta} F^{\alpha \beta} \, , 
\end{equation}
and $T_{e}^{\mu \nu}$ are the photon and electron energy momentum tensors, respectively, with the precise form of $T_{e}^{\mu \nu}$ dependent on how the charge carriers are modelled, e.g. a fluid model as in \cite{Macedo:1983wcr} or as a fundamental interaction of fermionic fields. In vacuum, (i.e., in the absence of charge carriers) it is easy to show that $\partial_{\mu } T^{\mu \nu}_{\gamma} =0   $, since $\partial_{\mu } F^{\mu \nu} = 0 $ in the absence of currents. However, when currents are present, $T^{\mu \nu}_\gamma$ and $T^{\mu \nu}_e$ are not separately conserved, and so energy-momentum conservation requires that \textit{both} terms be taken into account to ensure conservation of the full $T^{\mu \nu}$. 

In turn, this means that for (gravitational) gauge invariance to hold, one must consider the effects of both electromagnetic and matter fluctuations in order to obtain gauge-invariant results from the interaction vertex \eqref{eq:Lint}. Physically, this means that fluctuations in the electromagnetic field must be accompanied by fluctuations of charge carriers as well, and in turn, we expect that the presence of a magnetic field implies the presence of currents to generate it, and as we shall see, currents lead to additional electromagnetic fluctuations in the presence of gravitational waves. Hence, in some sense, the results we are about to present, show that  fluctuations from the Gertsenshtein effect must be accompanied by physical effects of charge carrier fluctuations too in order for gauge invariance to hold. Clearly, by working in TT gauge in which $h_{ij}$ are the only-non vanishing components, and $h^{i}_{\, \, \, i} =0 $, it follows that if the matter is isotropic, such that $T_{e}^{ij} \propto \delta^{ij}$, then $T^{\mu \nu}_e h_{\mu \nu} = 0$. However, this will not hold if one introduces an anisotropy in the background matter. One could see this explicitly in, e.g., a fluid model where $T_e^{\mu \nu} = (\rho + p)v^\mu v^\nu + p \eta^{\mu \nu}$, where if no currents are present, such that $v^\mu = (1,\textbf{0})$, then in the rest-frame of the plasma, $T^{ij_e } \propto \delta^{ij}$. By contrast, if the fluid is endowed with a finite velocity component, $v^i \neq  0$, then $T^{\mu \nu}_e h_{\mu \nu} = v^i v^j h_{i j}(\rho + p)$, inducing a direct coupling between the gravitational waves and the fluid velocity.

We now go on to demonstrate the explicit presence of secondary current fluctuations through a direct calculation in collisionless plasmas using kinetic theory. 

 \section{Gravitational waves in plasmas with currents} 

Let us consider the mixing between gravitational waves and photons in a background of charge carriers/plasma the absence of  a magnetic field. We know that we can write the wave equation for the electric field as
\begin{equation}
      \nabla^2 \textbf{E} - \nabla (\nabla \cdot \textbf{E}) - \ddot{\textbf{E}} = \dot{\textbf{J}}_{\rm ind } + \dot{\textbf{J}}_{\rm free}\,
\end{equation}
where 
\begin{equation}
    \textbf{J}_{\rm ind}(x) = \int d^4 x' \boldsymbol{\sigma}(x,x') \cdot \textbf{E}(x')
\end{equation}
is the induced current, $\boldsymbol{\sigma}_{ij}$ is the conductivity tensor (i.e. the response function),  and $\textbf{J}_{\rm free}$ is the free current, which can be induced by the motion of charges caused by a gravitational wave, which is described by the (collisionless) Boltzmann equation \cite{Macedo:1983wcr}
\begin{equation}
 k^\mu \partial_\mu f  + \left( e F^{\mu \nu} k_\nu - \Gamma^{\mu}_{\rho \sigma} k^\rho k^\sigma \right) \frac{\partial f}{\partial k^\mu} = 0  \, , 
\end{equation}
where $f$ is the phase space density of charge carriers. We can then expand all quantities as perturbations which are all $\mathcal{O}(h)$:
\begin{equation}
 f  =    \bar{f} + \delta f, \qquad g_{\mu \nu} = \eta_{\mu \nu} + h_{\mu \nu}, \qquad F_{\mu \nu} = \bar{F}_{\mu \nu}  + \delta F_{\mu \nu}, \qquad 
\end{equation}
where $\bar{f}$ is the background, unperturbed phase-space density of electrons, and $\bar{F}_{\mu \nu}$ is the background field strength tensor which in order to isolate the effect we are interested, we neglect it and set $\bar{F}_{\mu \nu} = 0$, but it could be included straightforwardly and would correspond to magnetised effects in the plasma conductivity and the usual Gertsenshtein current \cite{Berlin:2021txa} proportional to the background magnetic field. We then further decompose 
\begin{equation}
    \delta f = \delta f_{\rm ind} +  \delta f_{\rm free} \, ,
\end{equation}
where $\delta f_{\rm ind}$ captures fluctuations due to the induced electric field and $\delta f_{\rm free} $ captures perturbations due to the direct minimal coupling of the gravitational waves to charge carriers. This means that we have two sets of equations \begin{align}
 & k^\mu \partial_\mu \delta f_{\rm ind}  +  e \delta F^{\mu \nu} k_\nu   \frac{\partial \bar{f}}{\partial k^\mu} = 0  \, , \\
 & k^\mu \partial_\mu \delta f_{\rm free} - \Gamma^{\mu}_{\rho \sigma} k^\rho k^\sigma  \frac{\partial \bar{f}}{\partial k^\mu} = 0. \label{eq:free}
\end{align}
The first of these equations gives the usual expression for the response function of a magnetised plasma, which is well-know. The second shows the purely gravitational response of the particles due to minimal coupling. Indeed, the characteristics of the second equation give precisely the geodesics, i.e. $dx^\mu/d\lambda = k^\mu$ and $dk^\mu/d\lambda = - \Gamma^{\mu}_{\rho \sigma} k^\rho k^\sigma   = 0 $ which implies $d^2 x^\mu/d\lambda^2 + \Gamma^{\mu}_{\rho \sigma} d x^\rho/d\lambda d x^\sigma/d\lambda  = 0$. We are interested in the free current generated by the gravitational waves, which is given by
\begin{equation}
    J^\mu_{\rm free} = e \int dv^4 \,  v^\mu \delta f_{\rm free}. 
\end{equation}
where $v^\mu = (1 , \textbf{v})$ with $\textbf{v} = \textbf{k}/\omega$ is a timelike  vector field. We can then explicitly write Eq.~\eqref{eq:free} as
\begin{equation}
    \partial_t \delta f_{\rm free}  + \textbf{v}\cdot \nabla_\textbf{x} \delta f_{\rm free} =    \Gamma^{\mu}_{\rho \sigma} v^\rho v^\sigma  \frac{\partial \bar{f}}{\partial v^\mu} .
\end{equation}
Let us now multiply this by $v^\mu$ and integrate over $d^4 v$ to get a first moment of this equation
\begin{equation}
 \partial_t J_{\rm free}^\nu + \mathcal{O}(\textbf{v}^2) =
   e \int d^4 v \, \,v^\nu \Gamma^{\mu}_{\rho \sigma} v^\rho v^\sigma  \frac{\partial \bar{f}}{\partial v^\mu} \, , 
\end{equation}
where we have used the standard argument that for cold/non-relativistic plasmas we drop the second term on the left-hand side as it is quadratic in $\textbf{v}$. We can then integrate the right-hand side by parts with respect to $v^\mu$ to get 
\begin{equation}
 \partial_t J_{\rm free}^\nu  = - e
   \int d^4 v \, \,  \left[ \Gamma^{\nu}_{\rho \sigma} v^\rho v^\sigma    + 2 v^\nu \Gamma^{\mu }_{\mu \sigma}  v^\sigma\right] \bar{f} \, . 
\end{equation}
We can then use the explicit form of the Cristoffel symbols, which gives 
\begin{equation}
     \partial_t J_{\rm free}^\nu =  -  e \int d^4 v  
   \left(  \partial_{\rho} h^{\nu}_{\, \, \sigma} -\frac{1}{2}  \partial^{\nu} h_{\rho \sigma}  \right)v^\rho v^\sigma  \bar{f} -   e \int d^4 v  
    \partial_{\sigma} h_{
\mu}^{\, \, \mu}   v^\sigma   v^\nu \bar{f}  \, . 
\end{equation}
Now, if we work in the TT gauge, the second integral vanishes, since $ h_{
\mu}^{\, \, \mu}  =0 $, and since $h_{i 0} = h_{00} = 0$,  we have
\begin{equation}\label{eq:JFree}
     \partial_t \textbf{J}_{\rm free}^i =  -  e \left(  \partial_{\rho} h_{\, \, \sigma}^{i} - \frac{1}{2}\partial^{i} h_{\rho \sigma}  \right) \int d^4 v  
  \, v^\rho v^\sigma  \bar{f}(v) \, . 
\end{equation}
Let us therefore define a tensor integral object
\begin{equation}
    I^{\rho \sigma } \equiv \int d^4 v  
  \, v^\rho v^\sigma  \bar{f}(v) \, ,
\end{equation}
which appears on the right-hand side of \eqref{eq:JFree}.
If the plasma is \textit{isotropic}, such that
\begin{equation}
    \bar{f}(v_\mu) = f_{\rm iso}(v^2)
\end{equation}
there are no vectorial quantities which can appear in $I^{\rho \sigma }$, in which case it follows that $I^{\rho \sigma} \propto \eta^{\rho \sigma}$. However, we see by the TT conditions $\partial^\nu h_{\nu \mu} = 0$, that contraction with both the metric terms in Eq.~\eqref{eq:JFree} gives zero so that
\begin{equation}
      \partial_t \textbf{J}_{\rm free}^i  =  0 \, ,
\end{equation}
for an isotropic plasma. This would appear to be consistent with conclusions in \cite{Garg:2023yaw} which also finds that for transverse gravitational waves, no production of electromagnetic perturbations occurs in isotropic cold plasmas. 

Now let us suppose that there is a current flowing in the plasma, so that it is \textit{anisotropic}, with the background phase space density taking the form
\begin{equation}\label{eq:delta}
       \bar{f}(v_\mu)   = f_0(v) \delta^{(4)}(v - \bar{v}) \, .
\end{equation}
Upon Fourier transforming and assuming plane gravitational waves with frequency $\omega$ and 4-momentum $k_\mu$, we find
\begin{equation}\label{eq:CurrentFinal}
      \omega \textbf{J}_{\rm free}^i  = e \left(k_\rho h^{i}_{\, \, j}   \bar{v}^\rho \bar{\textbf{v}}^j - \frac{1}{2}\textbf{k}^i \bar{\textbf{v}}_l h_{l m} \bar{\textbf{v}}_m \right) f_0 \, . 
\end{equation}
We see immediately, that if the current is parallel to the gravitational wave, then everything vanishes by the TT conditions. Assuming that the background plasma is non-relativistic such that the second term quadratic in $v_i$ can be neglected (with $\bar{v}^0 \simeq 1$) and $k_0 = \omega$, we have
\begin{equation}\label{eq:Induced}
       \textbf{J}_{\rm free}^i  \simeq    h^{i}_{\, \, j}    \bar{\textbf{v}}^j f_0  \equiv  h^{i}_{\, \, j} \textbf{J}^j_{\rm bck} \, ,
\end{equation}
where $\textbf{J}^j_{\rm bck} \equiv e \int d^3\textbf{v} \, \bar{\textbf{v}}^j \,\bar{f}(v) = e \bar{ \textbf{v}} f_0(\bar{\textbf{v}} ) $ is the background current, and in performing the integral, we used the relation \eqref{eq:delta}. 

Eq.~\eqref{eq:Induced} is the central result of this paper, which shows that a combination of gravitational waves and background currents can induce a secondary AC current $ \textbf{J}_{\rm free}^i$ in the plasma. It will clearly be of great interest to investigate the phenomenological consequences of this effect in future work.

\section{Summary}

In this short calculation, we have demonstrated how gravitational waves propagating in plasmas in which a background current is present can generate secondary electromagnetic currents. By contrast, we noted that for isotropic plasmas no such secondary charge/current fluctuations occur. As a result, a plasma containing a DC current will experience a secondary AC current due to the presence of gravitational waves. This makes sense from the point of view of trying to construct a linear relation between a secondary current $j^{\mu}$ and the tensorial quantity $h_{\mu \nu}$. Clearly, any such correspondence would require some directed vectorial background quantity to facilitate the relation and allow appropriate contraction of indices. For the Gertsenshtein effect, this is facilitated by the background magnetic field (see, e.g., \cite{Berlin:2021txa}), whilst here, the current plays this role, with both breaking isotropy. Perhaps microphysical arguments can also be made on the basis of symmetries, e.g. helicity conservation, or similar, though we shall not delve into such arguments here. 

Clearly it will be interesting to investigate the phenomenological consequences of such currents as a route to the detection of gravitational waves in astrophysical plasmas, but also via analogous currents in conductors in laboratory setups (see the note at the end of this manuscript). 
\\\\
\noindent \textbf{Acknowledgments.} We thank Sebastian Ellis, Pete Millington and Alessandro Principi for useful discussions and acknowledge support form the Science and Technology
Facilities Council (STFC) [Grant No. ST/X00077X/1]. 
\\\\

\noindent \textbf{Further Note.} Prior to release, we shared the calculation presented in this work with Sebastian Ellis, who kindly made us aware of the preparation of related results which explore similar effects arising from current flows in conductors. We have released our results concurrently with that work.

\bibliography{bibliography.bib}{}
\bibliographystyle{JHEP}

\end{document}